# Isotopic and velocity distributions of $_{83}$Bi produced in charge-pickup reactions of $^{208}_{82}Pb$ at 1 $A$ GeV

A. Kelić [a], K.-H. Schmidt [a], T. Enqvist [a,#], A. Boudard [b], P. Armbruster [a], J. Benlliure [c], M. Bernas [d], S. Czajkowski [e], R. Legrain [e,§], S. Leray [b], B. Mustapha [d, &], M. Pravikoff [e], F. Rejmund [d], C. Stéphan [d], J. Taïeb [b], L. Tassan-Got [d], C. Volant [b], W. Wlazło [b, f]

[a] *GSI, Planckstraße 1, D-64291 Darmstadt, Germany*
[b] *DAPNIA/SPhN CEA/Saclay, F-91191 Gif-sur-Yvette, France*
[c] *University of Santiago de Compostela, E-15706 Santiago de Compostela, Spain*
[d] *IPN Orsay, IN2P3, F-91406 Orsay, France*
[e] *CENBG, IN2P3, F-33175 Gradignan, France*
[f] *Jagiellonian University, Institute of Physics, ul. Reymonta 4, P-30-059 Kraków, Poland*

**Abstract:** Isotopically resolved cross sections and velocity distributions have been measured in charge-pickup reactions of 1 $A$ GeV $^{208}$Pb with proton, deuterium and titanium target. The total and partial charge-pickup cross sections in the reactions $^{208}$Pb + $^1$H and $^{208}$Pb + $^2$H are measured to be the same in the limits of the error bars. A weak increase in the total charge-pickup cross section is seen in the reaction of $^{208}$Pb with the titanium target. The measured velocity distributions show different contributions – quasi-elastic scattering and Δ-resonance excitation - to the charge-pickup production. Data on total and partial charge-pickup cross sections from these three reactions are compared with other existing data and also with model calculations based on the coupling of different intra-nuclear cascade codes and an evaporation code.

## 1. Introduction

In nuclear charge-pickup reactions at projectile energies well above the Fermi energy there are two processes mostly responsible for the increase of the nuclear charge of the projectile [1]: One is a quasi-elastic collision between a target proton and a projectile neutron where the proton takes over the total kinetic energy of the neutron ending up in the phase volume of the projectile-like fragment. The other mechanism is the excitation of a target or a projectile nucleon into the Δ(1232)-resonance state and its subsequent decay. Typical transfer reactions observed at lower energies [2] where a proton is transferred through the nuclear overlap zone are excluded at relativistic energies because of the non-overlapping Fermi spheres of projectile and target. This is also confirmed by the non-observation of fragments with masses higher than the mass of the projectile [3].

In previous works special attention was devoted to the charge-exchange reactions, where the nuclear charge of the projectile increases or decreases by one unit but where no mass loss occurs. As at relativistic energies charge-exchange reactions involve the formation of Δ–particles and pions, they can be used as a tool for studying the in-medium behaviour of these particles [4]. These reactions can also give some insight into the neutron density distribution in the nucleus [5].

---
[#] Present address: *University of Jyväskylä, FIN-40351 Jyväskylä, Finland.*
[§] Deceased.
[&] Present address: *Argonne National Laboratory, 9700 South Cass Avenue, Building 203, Argonne, IL 60439, USA.*

The most comprehensive study of charge-exchange reactions involving relativistic heavy ions (up to $^{40}$Ar) has been performed at the accelerator SATURNE at Saclay as a part of a research program dedicated to the investigation of the nuclear response to spin-isospin excitations; for an overview see reference [1]. An interesting finding from the measured inclusive momentum spectra of the reaction products was a shift of the position and of the strength of the Δ-resonance peak in heavier targets as compared to the proton target [1]. Half of this shift is explained by the Fermi motion of the Δ–particles and the nucleons in the nuclear mean field [6, 7], while effects such as the strongly attractive interaction between Δ-particle nucleon-hole states in the spin-longitudinal channel [4] and the interplay between Δ-excitation in the target and in the projectile [8] could contribute to another half.

For projectiles heavier than $^{40}$Ar, most of the published data concern only total charge-pickup cross sections; see [9, 10, 11] and references there in. Guoxiao et al. tried to establish the systematic dependence of the total charge-pickup cross section on projectile and target size [12]. Using the data measured for a wide range of projectile masses, from $^{12}$C to $^{197}$Au, they found evidence for a square dependence of the total charge-pickup cross section on the projectile mass. To explain this steep dependence, they suggested the presence of coherent processes in the charge-pickup reactions. Later, using complete measured isotopic distributions of charge-pickup products, Sümmerer et al. [13, 14] showed that the scaling of the total charge-pickup cross section with the square of the projectile mass is mostly due to the evaporation of charged particles from the prefragments. Therefore, they concluded that the total charge-pickup cross sections are not a sensitive tool for studying the basic nucleon-nucleon processes involved in charge-pickup reactions.

In the present work, we report on the isotopically resolved charge-pickup cross sections for the production of $_{83}$Bi in the interactions of relativistic, 1 A GeV, $^{208}_{82}$Pb projectiles with proton, deuterium and titanium target. The measurements were performed at GSI-Darmstadt using the full advantage of relativistic collisions in inverse kinematics. These data represent a part of a comprehensive study of fragment formation in a neutron-generating target for accelerator-driven systems (ADS) [15, 16, 17].

The present paper is organised in the following way: In section 2 we give a short overview on the experiment and the data analysis. Section 3 is dedicated to the experimental results concerning total and partial charge-pickup cross sections for bismuth production as well as the velocity distribution of each bismuth isotope. In section 4 we compare the data from this work with other published data on total and partial charge-pickup cross sections. Comparisons of the measured data with the results of calculations performed with two intra-nuclear cascade models coupled to an evaporation/fission code are presented in section 5. Conclusions are presented in section 6.

## 2. Experiment and data analysis

In order to obtain high-precision experimental data on mass, atomic number and momentum distributions of reaction residues we made use of a high-resolution magnetic spectrometer. The magnetic rigidity of each residue was measured with high precision just by determining the deflection in a magnetic dipole field. This gives very precise information on the residue's longitudinal momentum and hence on its velocity, once the residue was identified in mass and atomic number. The experimental method and the data-analysis procedure concerning these measurements have been described in detail in reference [16]. Here only a short overview will be given.

## 2.1. Experiment

The experiments were performed at GSI-Darmstadt, Germany. The primary beam of $^{208}$Pb at an energy of 1 $A$ GeV was delivered by the heavy-ion synchrotron SIS. The fragment separator FRS [18] and the associated detector equipment (Figure 1) were used in order to separate and to identify the reaction products. The FRS is a two-stage magnetic spectrometer with a dispersive intermediate image plane (S2) and an achromatic final image plane (S4). The momentum acceptance is 3 %, and the angular acceptance is about 15 mrad around the beam axis. Two position–sensitive plastic scintillators [19], with a thickness of 5 mm and dimensions of (218 x 80) mm$^2$ and (200 x 80) mm$^2$, were placed at S2 and S4, respectively. These detectors provided the information on magnetic rigidity ($B\rho$) and the time-of-flight ($TOF$) of each reaction product.

In order to achieve the necessary nuclear-charge resolution, the reduction of magnetic rigidity due to the energy loss in a profiled aluminium degrader [20] (thickness 5236 mg/cm$^2$, installed at S2) was used in combination with a multi-sample ionisation chamber MUSIC [21].

The beam-current monitor SEETRAM [22, 23] was continuously in use in order to measure and to check the primary-beam intensity.

The proton and the deuteron targets were realised as liquid targets enclosed between thin titanium foils of a total thickness of 36.3 mg/cm$^2$ [24]. The thicknesses of the proton and deuteron targets were measured to be (87.3 ± 2.2) mg/cm$^2$ [25] and (206 ± 6) mg/cm$^2$ [17], respectively. To maximise the number of bare ions passing through the FRS, niobium stripper foils of thicknesses 60 mg/cm$^2$ and 106 mg/cm$^2$ were set behind the target and behind the degrader at S2, respectively. In order to subtract the contribution from different layers of matter (the beam window, the beam monitor, the target windows and the stripper foil) to the measured production rate, the measurements were repeated replacing the proton and deuteron targets by an empty target container.

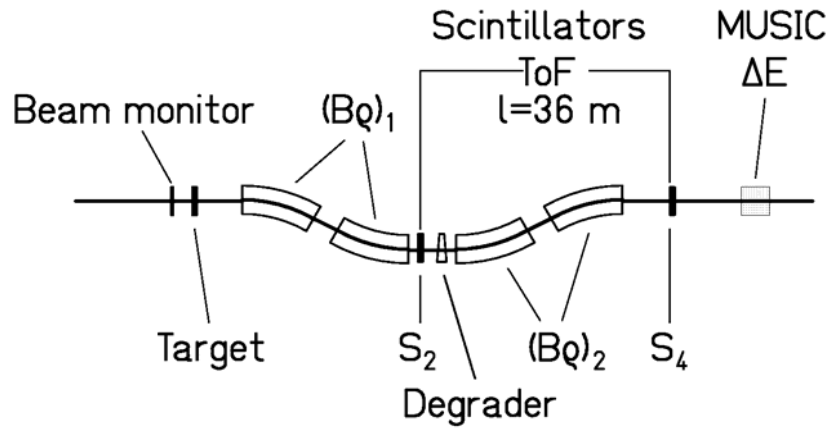

**Figure 1.** Schematic drawing of the fragment separator FRS with the associated detector equipment. For more details see text.

## 2.2. Data analysis

The data analysis was based on the reconstruction of the full velocity distribution of each isotope. For an unambiguous identification of bismuth isotopes, only the completely stripped ions passing all along the FRS were considered in the data analysis by applying the method of a two-fold energy-loss measurement [16].

Selecting only events fulfilling the above condition, the nuclear-charge and mass calibrations were performed in the following way:

- For each magnetic-field setting, the events corresponding to nuclear charge equal to 83 were selected from the two-dimensional spectra of the position at the final focal plane (S4) versus the position at the intermediate focal plane (S2), Figure 2.
- After selecting the charge, the mass was determined from the two-dimensional spectra of the time-of-flight in the second half of the FRS versus the position at S2. Because of the limited momentum acceptance of the FRS, for a given magnetic-field setting only few bismuth isotopes are transmitted, see Figure 2. In order to overcome this, the data obtained with different magnetic-field settings by scanning over the momentum distribution were combined.

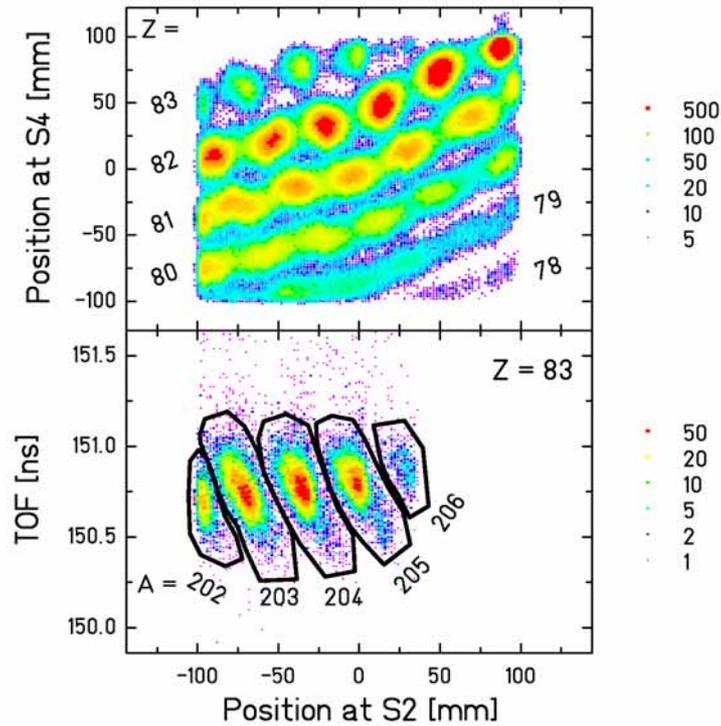

**Figure 2**. Charge and mass identification of bismuth products measured in the reaction $^{208}$Pb + $^{1}$H. Upper part: Two-dimensional spectrum of the position at S4 versus the position at S2. Bands corresponding to different charges are marked according to Z. Lower part: Two-dimensional spectrum of the time-of-flight in the second half of the FRS versus the position at S2 for the events corresponding to Z = 83; the different windows correspond to different bismuth isotopes (A = 202, 203, 204, 205, 206). Both figures show data measured for only one setting of the magnetic fields. Therefore, due to the limited momentum acceptance of the FRS, the isotopes $^{202}$Bi and $^{206}$Bi are only partly transmitted.

## 2.3. Determination of cross sections and associated uncertainties

After the identification of A and Z, the longitudinal momentum $p$ of each fragment was recalculated from the following equation:

$$p = \frac{e}{c} \cdot Z \cdot B\rho , \qquad (1)$$

Here, $B\rho$ is the value of the fragment magnetic rigidity in the first half of the FRS, c the velocity of light and -e the charge of the electron. In this way, the resolution with which the momentum is obtained is given only by the resolution in $B\rho$ (after identification, Z is an integer number), $\Delta B\rho/B\rho \approx 5 \cdot 10^{-4}$, and is improved by one order of magnitude as compared to the resolution obtained from the TOF measurement. From the momentum, the velocity υ of each fragment was obtained as:

$$\beta \cdot \gamma = \frac{p}{M(A,Z) \cdot c}, \quad \upsilon = \beta \cdot c \tag{2}$$

where, γ is the relativistic Lorentz factor and $M(A,Z)$ the mass of the nucleus $(A,Z)$. This velocity was then transformed into the reference frame of the primary beam in the middle of the target by Lorentz transformation, taking into account the appropriate energy losses of both the projectile and the fragment. Thus, for each isotope the velocity distribution was obtained after correcting for the dead time of the data-acquisition system, for losses due to secondary reactions in the degrader and in the scintillator at S2, for losses coming from the rejections of the incompletely stripped ions from the data analysis, and also normalising to the number of counts in the beam-current monitor. The production cross section of a specific isotope was obtained from its production rate given by the surface below the peak in the corresponding velocity distribution corrected for the contribution from the target container, and normalised to the number of target atoms per area. The production rate in the target container was about 4% of the production rate in the full target. Finally, one needs to correct the measured data for the limited momentum and angular acceptance of the FRS. In the present work, the limited momentum acceptance of the fragment separator is not crucial, since, as already mentioned, the momentum distributions of all fragments were fully measured by superposing the measurements with different settings of the magnetic fields. Moreover, the angular range of produced bismuth isotopes was fully covered by the angular acceptance of the fragment separator [26], resulting in the angular acceptance of 100%. All the applied corrections are explained in more detail in reference [16].

The production cross sections for bismuth isotopes were also extracted from the measurements performed with the empty target container. The empty target container consists mostly of four titanium foils. Additionally, thin Mylar foils coated with a very thin aluminium layer represent the thermal isolation of the cryogenic target. In deducing the cross sections from the measurements with the empty target container one has to take into account all layers of matter present in the target area. In Table 1, these layers are listed together with the corresponding thicknesses and numbers of atoms per area (for Mylar, this number corresponds to the total number of atoms and not to the number of molecules). As titanium contributes mostly to the number of target nuclei, ranging from hydrogen to niobium, of the materials present in the target area (see the fourth column of Table 1), from now on we will refer to this target as "Ti", although one should keep in mind its complex composition.

**Table 1.** The list of the layers of matter present in the target area during the measurement with the empty target container.

| Layer | Material | Thickness [mg / cm$^2$] | N$^o$ atoms·10$^{20}$ / cm$^2$ |
|---|---|---|---|
| Ti beam window | Ti | 4.5 | 0.57 |
| SEETRAM | Al | 8.9 | 1.99 |
| Ti target windows | Ti | 36.3 | 4.57 |
| Nb-stripper | Nb | 60.0 | 3.89 |
| Mylar | C$_5$H$_4$O$_2$ | 8.3 | 0.52 |
| Aluminium | Al | 0.1 | 0.02 |

The systematic uncertainty of the production cross sections of bismuth isotopes, due to the uncertainties of all mentioned corrections, was estimated to be 9 %.

## 3. Results
### 3.1 Total charge-pickup cross sections

The measured total charge-pickup cross sections for the production of $_{83}$Bi in the reactions of 1 $A$ GeV $^{208}$Pb in the proton-, deuteron-, and "Ti"-target are given in Table 2. These cross sections show a weak dependence on the size of the target nucleus, increasing by less than a factor of two when going from the proton to the "Ti" target. Because of the strong absorption in nuclei, one expects these reactions to be very peripheral and, therefore, their cross sections not much influenced by the target size. For similar systems, the same behaviour has been seen in some other experiments [9, 10, 11].

**Table 2**. Measured and calculated total charge-pickup cross sections for bismuth production in the reactions of $^{208}$Pb (1 $A$ GeV) with different targets. The calculations are described in section 5.

|  | $^{208}$Pb + $^{1}$H $\sigma$ [mb] | $^{208}$Pb + $^{2}$H $\sigma$ [mb] | $^{208}$Pb + "Ti" $\sigma$ [mb] |
|---|---|---|---|
| Experiment | 30 ± 6 | 26 ± 6 | 45 ± 9 |
| INCL4+ABLA | 36 | 31 | — |
| ISABEL+ABLA | 34 | 33 | 51 |

In Figure 3, the total charge-pickup cross sections are shown as a function of the target mass for the following projectiles: 1 $A$ GeV $^{208}$Pb from the present work, 0.915 $A$ GeV $^{197}$Au [11], 1.2 $A$ GeV $^{197}$Au [9] and 1 $A$ GeV $^{208}$Pb [27].

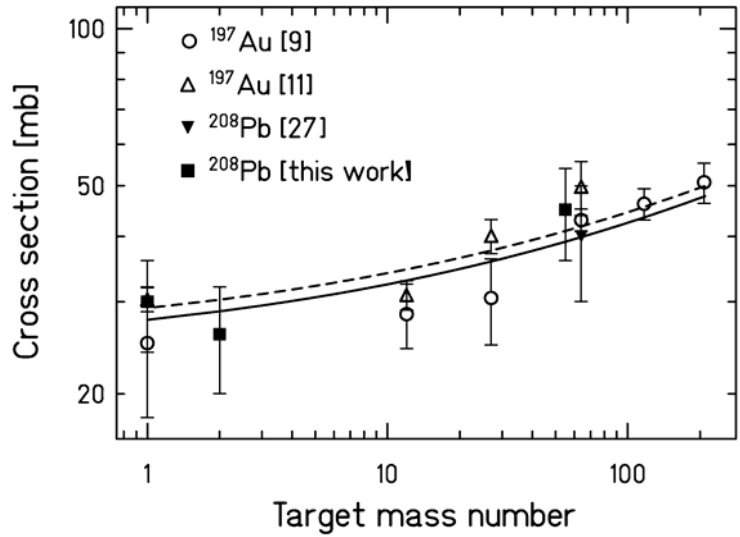

**Figure 3.** Total projectile charge-pickup cross section as a function of the target mass: open triangles - $^{197}$Au at 0.915 $A$ GeV [11], full squares - $^{208}$Pb at 1 $A$ GeV from the present work (note: for the reaction $^{208}$Pb + "Ti" the data point is shown for the weighted mean mass of all materials composing this target), open dots - $^{197}$Au at 1.2 $A$ GeV [9], and full triangles - $^{208}$Pb at 1 $A$ GeV [27]. The full (Au projectile) and dashed (Pb projectile) lines are obtained by using the following relation: $\sigma = 0.12 \cdot (A_p^{1/3} + A_t^{1/3}) \cdot A_p^{2/3}$, where $A_p$ and $A_t$ are the values of projectile and target mass, respectively.

Only the data measured with lead and gold projectiles are included because, on the chart of the nuclides, these projectiles are both quite far away from the evaporation corridor [28]. As a consequence, the deexcitation process of charge-pickup prefragments formed in these reactions is mostly governed by neutron evaporation. This is not true for lighter projectiles situated close to the evaporation corridor, where the competition between neutron and proton evaporation reaches its asymptotic value quickly, and the final, observed total charge-pickup cross sections can differ significantly from the primary charge-pickup cross sections. On the other hand, in the case of projectiles heavier than lead, fission tends to deplete considerably the primary charge-pickup production.

Following ideas from Refs. [10, 13] we have parameterised the total charge-pickup cross sections shown in Figure 3. As the charge-pickup reactions are expected to be peripheral reactions, the cross section should be proportional to $A_p^{1/3} + A_t^{1/3}$, with $A_p$ and $A_t$ being the values of projectile and target mass, respectively. On the other hand, in a charge-pickup reaction at least one scattered nucleon from a (p, n) or (n, p) elementary reaction must be reabsorbed by the projectile. As the probability for re-absorption should be proportional to the projectile surface, one expects that the total charge-pickup cross section should scale approximately as $(A_p^{1/3} + A_t^{1/3}) \cdot A_p^{2/3}$.

Following these arguments, the total charge-pickup cross sections for lead and gold projectiles as a function of the target mass were parameterised as: $\sigma = 0.12$ mb$\cdot(A_p^{1/3}+A_t^{1/3})\cdot A_p^{2/3}$. The factor 0.12 is roughly determined in order to scale the calculated cross sections to the measured ones. Results of this parameterisation are shown in Figure 3 with full and dashed lines for gold and lead projectiles, respectively. The agreement of this simple parameterisation with the experimental data is very good.

### 3.2 Isotopic production cross sections

The isotopic distributions of bismuth isotopes measured in the present work are shown in Figure 4. In case of the "Ti" target, as the measured statistics was much lower than in case of the $^1$H- or the $^2$H-target, it was not possible to extract cross sections for those bismuth isotopes that were only partly transmitted in two consecutive magnetic-field settings.

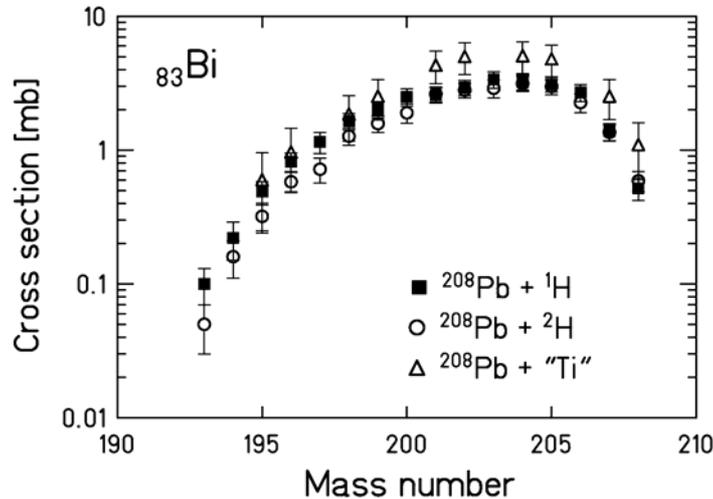

**Figure 4**. Production cross sections of bismuth isotopes from the 1 $A$ GeV reactions $^{208}$Pb + $^1$H - full squares, $^{208}$Pb + $^2$H - open dots, and $^{208}$Pb + "Ti" - open triangles. Error bars correspond to total uncertainties, statistical plus systematic.

Going from the proton and the deuteron target to the "Ti" target, the increase in the total charge-pickup cross section is mostly taken by the more neutron-rich isotopes (above $^{201}$Bi). The most neutron-deficient bismuth isotopes are produced with quite similar cross sections, regardless of the target.

The partial production cross sections for bismuth isotopes from the reactions of $^{208}$Pb with protons, deuterons and the "Ti" target are given in Table 3.

**Table 3.** Measured partial charge-pickup cross sections from the reactions $^{208}$Pb + $^{1}$H, $^{208}$Pb + $^{2}$H and $^{208}$Pb + "Ti" target at 1 A GeV. Only statistical uncertainties are given. The systematic uncertainty amounts to 9%.

| Mass number | $^{208}$Pb + $^{1}$H $\sigma$ [mb] | $^{208}$Pb + $^{2}$H $\sigma$ [mb] | $^{208}$Pb + "Ti" $\sigma$ [mb] |
|---|---|---|---|
| 193 | 0.10 ± 0.02 | 0.05 ±0.02 | |
| 194 | 0.22 ±0.05 | 0.16 ±0.04 | |
| 195 | 0.49 ±0.05 | 0.32 ±0.04 | 0.60 ±0.31 |
| 196 | 0.82 ±0.06 | 0.58 ±0.05 | 0.97 ±0.39 |
| 197 | 1.15 ±0.11 | 0.72 ±0.09 | |
| 198 | 1.64 ±0.08 | 1.27 ±0.07 | 1.86 ±0.52 |
| 199 | 2.02 ±0.10 | 1.58 ±0.08 | 2.53 ±0.62 |
| 200 | 2.50 ±0.15 | 1.90 ±0.14 | |
| 201 | 2.60 ±0.11 | 2.63 ±0.10 | 4.32 ±0.80 |
| 202 | 2.94 ±0.12 | 2.81 ±0.11 | 5.01 ±0.88 |
| 203 | 3.38 ±0.18 | 2.90 ±0.18 | |
| 204 | 3.21 ±0.12 | 3.15 ±0.12 | 5.08 ±0.88 |
| 205 | 3.11 ±0.12 | 2.99 ±0.12 | 4.81 ±0.86 |
| 206 | 2.70 ±0.15 | 2.27 ±0.16 | |
| 207 | 1.38 ±0.08 | 1.36 ±0.07 | 2.53 ±0.61 |
| 208 | 0.52 ± 0.05 | 0.59 ±0.05 | 1.10 ±0.40 |

### 3.3 Velocity distributions

The velocity distributions of charge-exchange products display the different regions of nuclear excitation [1]: a peak at the velocity close to that of the beam corresponding to a quasi-elastic interaction and a broader peak at lower velocities caused by the excitation of a nucleon in the Δ(1232)-resonance state. As a magnetic spectrometer, the FRS allows one not only to measure the production cross sections but also the kinematical properties of the reaction residues with high resolution. This property of the FRS was of crucial importance in experiments such as the observation of deeply bound pionic states in nuclei [29], measurements of the momentum distribution of individual nucleons in halo nuclei [30, 31] or the investigation of the response of the spectator to the participant blast [32].

In order to improve the experimental velocity resolution, several corrections to the measured distributions had to be applied:

*-Target thickness and energy-loss straggling in the target -* If the target would be infinitely thin, the fragment velocity distribution (mean value and width) would be determined only by the reaction mechanism. Since a target has a finite thickness, the mean value and the width are modified by the energy loss of the beam (from the entrance to the place where the fragment is formed), by the energy loss of the fragment (from the place where the fragment is formed to the exit), and by the energy-loss straggling. The contribution of these effects to the width[a] of the velocity distributions can be represented as the convolution of a Gaussian function given by the energy-loss straggling and a square function given by the target thickness [33]. For each isotope, the widths of these two contributions were calculated with the code AMADEUS [20], and transformed into the reference frame of the projectile.

*- Momentum spread of the beam, size of the beam spot at the target position and position resolution of the scintillator -* The contribution of the first effect is reflected in the finite width of the momentum distribution of the primary beam, while the second effect results in the

---
[a] The mean value was already corrected for the energy losses of the projectile and the fragment, Section 2.3.

finite width of the position distribution at the intermediate image plane (S2). The position resolution of the scintillator contributes additionally to these widths. The width of the apparent momentum distribution containing the contributions of the above mentioned effects was deduced from the calibration measurement performed with the primary beam without the target, and its corresponding velocity value in the projectile frame was found to be FWHM = (0.0143 ± 0.0004) cm/ns. A Gaussian of that width was convoluted with the response function coming from the target thickness and the energy-loss straggling. The obtained response function was used to deconvolute, for each isotope, the measured velocity distribution represented in the reference frame of the projectile.

- *Higher-order ion-optical aberrations* - Second-order aberrations were corrected for by using sextupole magnets during the experiment placed in front and behind each dipole magnet. The contributions from higher order (>2) aberrations were neglected.

As an example, Figure 5 shows the velocity of $^{207}$Bi produced in the interactions of $^{208}$Pb with the proton target before (dashed histogram) and after (full histogram) the above-mentioned corrections. The contributions from the two different mechanisms, quasi-elastic scattering and $\Delta(1232)$-resonance formation, to the production are clearly visible in the corrected spectrum.

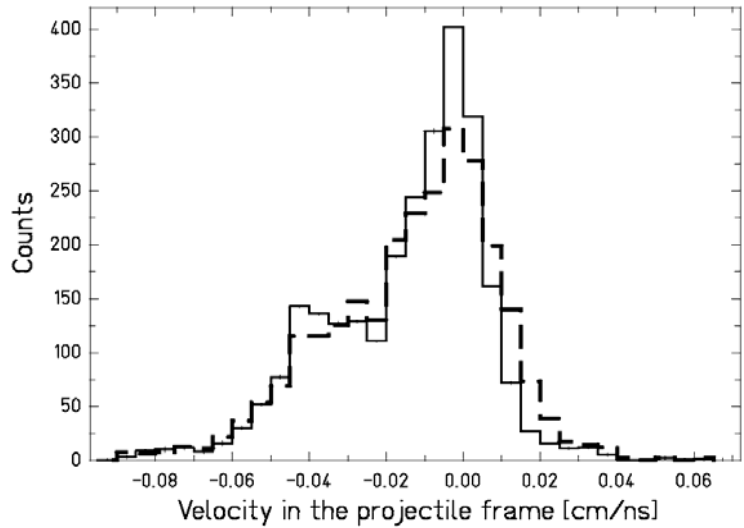

**Figure 5.** Velocity of $^{207}$Bi produced in the reaction $^{208}$Pb + $^{1}$H measured in the present work. The dashed histogram shows the raw velocity spectrum, while the full histogram shows the velocity after applying the deconvolution mentioned in the text.

The deconvoluted velocity distributions are shown in Figure 6 for several bismuth isotopes. For the most neutron-deficient isotopes, the velocity distributions are wide, and the above-mentioned corrections do not have any influence. In the case of the "Ti" target, the measured statistics was not high enough for the corrections to be applied. Therefore, in Figure 6 we do not show the velocity distributions of $^{208}$Bi and $^{207}$Bi measured in the "Ti" target. In the same figure, on the upper axis is shown the energy transfer in the laboratory frame, which was calculated from the measured velocities of the beam and the residue assuming that no mass loss has occurred. For $^{208}$Bi, events with negative energy transfer reflect the finite resolution of the experiment. In the cases $A < 208$ the two-body kinematics applied for calculating the energy transfer is not strictly valid, which additionally contributes to the events with apparent negative energy transfer.

The general behaviour of the velocity distributions, shown in Figure 6, is widening of the width and lowering of the mean value with decreasing mass of the residue. These effects are expected, as the lighter bismuth isotopes are likely to come from events involving higher excitation energy.

Looking in more details, one can see a clear appearance of two components for $^{208}$Bi, corresponding to quasi-elastic scattering and Δ–resonance excitation. Going from the proton to the deuteron target, the quasi-elastic contribution leading to the production of $^{208}$Bi decreases by a factor 1.4, which is obtained by a fit with two-Gaussians. The quasi-elastic component of a charge-pickup reaction can occur only on the target proton and not on the target neutron. Consequently, the quasi-elastic component is expected to be about two times smaller in the deuteron case. Distortion effects of the deuteron target could influence this ratio.

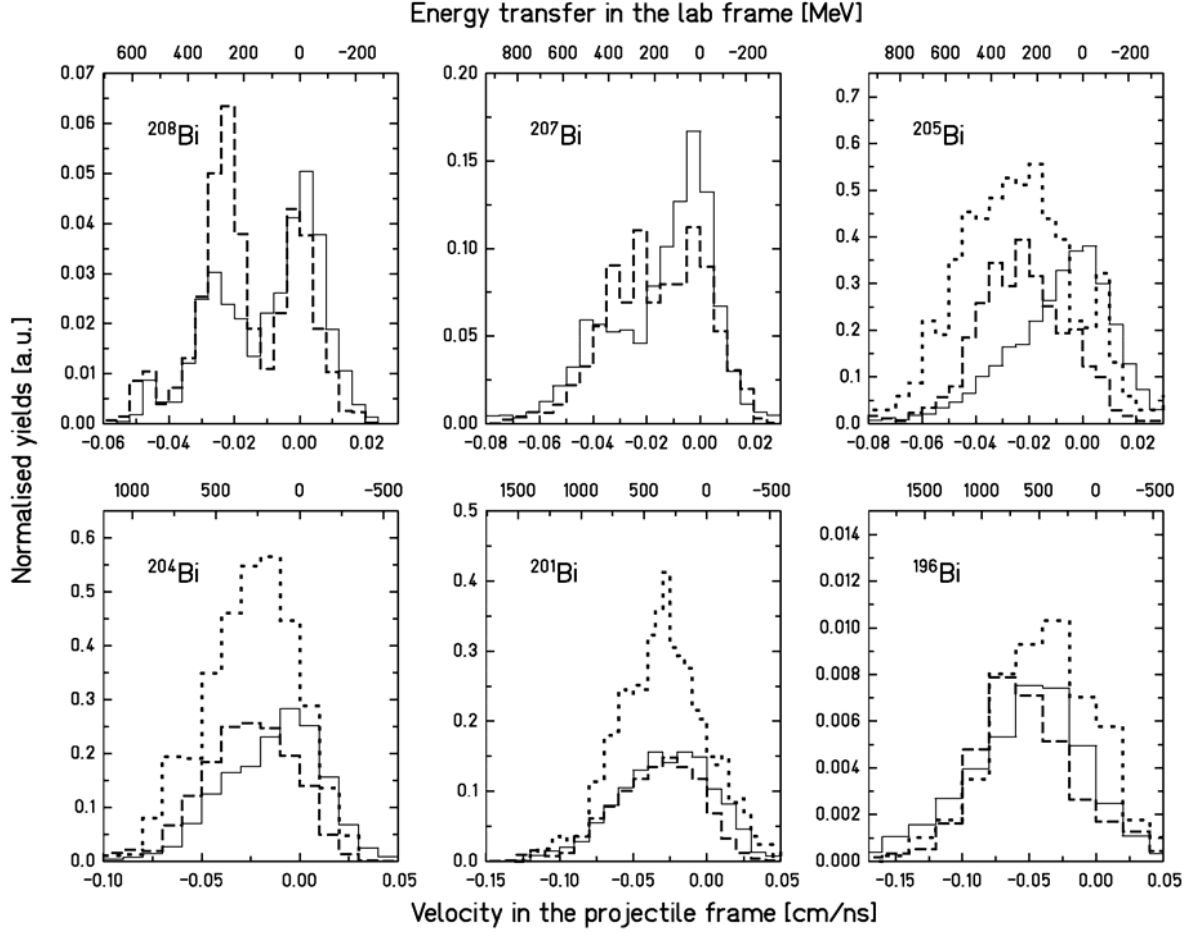

**Figure 6.** Longitudinal velocities (lower scale) of several bismuth isotopes produced in the interaction of 1 $A$ GeV lead with the proton- (full line), deuteron- (dashed line) and "Ti"-target (dotted line). The velocity distributions are normalised to the corresponding production cross sections. The upper scale represents the energy transfer in the laboratory frame.

On the other hand, the contribution of the Δ–resonance excitation to the formation of $^{208}$Bi is larger for the deuteron target than for the proton target. In case of the proton target a process leading to the production of $^{208}$Bi is p($^{208}$Pb, $^{208}$Bi)$\Delta^0$, while in case of the deuteron target, depending on the orientation of a deuteron with respect to the projectile, one can excite the Δ–resonance either on the target proton or on the target neutron via n($^{208}$Pb, $^{208}$Bi)$\Delta^-$ reaction. Considering that the isospin Clebsch-Gordan coefficient for neutrons is three times larger than for protons [8], one would expect the Δ-resonance component to be about two times stronger in case of the deuteron target as compared to the proton target. A distortion of the deuteron could also influence this ratio. From the fits to the measured velocity distributions of $^{208}$Bi, we obtain that the Δ–resonance contribution for the deuteron target is by a factor of 1.7 larger as compared to the proton target. From the same fit, we have obtained that the mean energy transfer corresponding to the Δ–resonance contribution is equal to (293

±12) MeV and (274±12) MeV for proton and deuteron target, respectively, which is in agreement with other data [4].

A third peak at the velocity of –0.047 cm/ns (corresponding to an energy transfer of ≈ 539 MeV) is also visible in the velocity distribution of $^{208}$Bi for both the proton and the deuteron target. Due to the low statistic inside this peak we do not discuss it further.

For lighter bismuth isotopes, the quasi-elastic component gradually disappears, and for the lightest isotopes only the contribution from the Δ–resonance excitation is present. An interesting finding is that for the isotopes $^{204-207}$Bi produced on the proton target, the quasi-elastic component is stronger as compared to the deuteron or the "titanium" target. For the lightest bismuth isotopes ($A$ < 204) the overall shape and mean value of the velocity distributions are very similar for all three targets.

We determined the mean values of the velocity distributions, calculated in the rest frame of the projectile, for all bismuth isotopes and determined the variation of this mean value with the mass loss. This dependence is shown in Figure 7. In the same figure, these mean values are compared with the empirical systematics of Morrissey [34]. The data and the systematics show the same general tendency, namely the decrease in the mean velocity with increasing mass loss, but the data show a larger reduction as compared to the systematics. The Morrissey systematics [34] was obtained from the analysis of fragmentation data, where the events with larger mass losses correspond to less peripheral collisions. In these cases, the target- and the projectile-nucleus penetrate each other more, and, therefore, the friction is increased. This then results in a larger loss of the projectile kinetic energy, and consequently in a larger reduction in the velocity of the projectile residue. On the other hand, in the charge-pickup process a larger part of the projectile kinetic energy can be spent on the formation of a Δ–resonance. Therefore, in this case, smaller mass losses can be connected with larger reduction in velocity than given by the Morrissey systematics.

The rather small velocity reduction in the proton-induced reaction for mass losses up to seven units reflects the particularly strong contribution of quasi-elastic scattering.

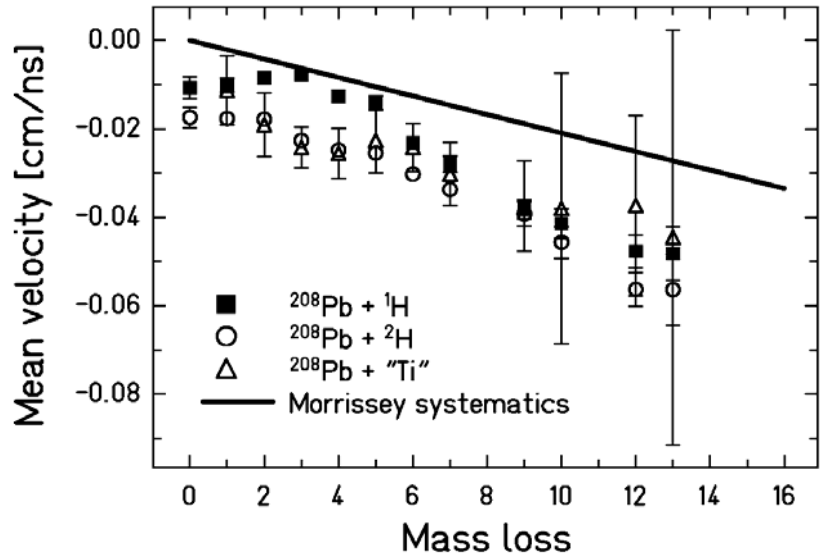

**Figure 7.** Mean values of measured longitudinal velocities of bismuth isotopes produced in the reactions $^{208}$Pb+$^{1}$H (full squares), $^{208}$Pb+$^{2}$H (open dots) and $^{208}$Pb+"Ti" (open triangles) at 1 $A$ GeV, and the empirical systematics of Morrissey [34] - full line. Velocities are shown in the rest frame of the $^{208}$Pb projectile.

## 4. Comparison with other data

Most published data from charge-pickup reactions at relativistic energies concern total cross sections (for example [9, 10, 11, 12] and references therein). The partial cross sections for a few isotopes were measured using the method of γ spectroscopy [35, 36, 37, 38]. To our knowledge, it is only from the measurements here at GSI (Refs. [13, 14, 25, 27, 39], and the present work) that full isotopic distributions of charge-pickup products are available.

Figure 8 shows the total charge-pickup cross section for the reaction $^{208}$Pb + $^{1}$H from the present work compared with the data from $^{197}$Au + $^{1}$H at 0.8 $A$ GeV [25], $^{197}$Au + $^{1}$H at energies below 1 $A$ GeV [11] and the same reaction at projectile energies above 1 $A$ GeV [9]. In order to see more clearly the effect of the projectile energy, only reactions involving similar projectiles and the same target are compared. For energies below ~ 2 $A$ GeV, the total charge-pickup cross section decreases rapidly with increasing projectile energy. On the other hand, for projectile energies between 2 and 10 $A$ GeV the total charge-pickup cross section seems to be independent of the energy involved in the reaction.

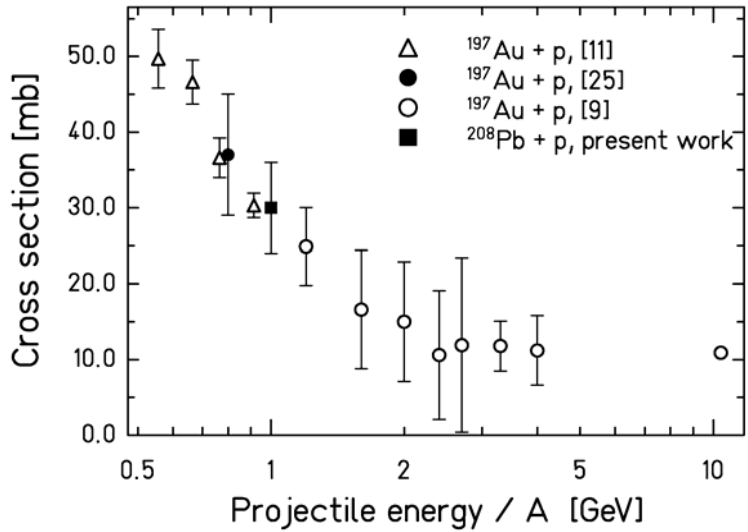

**Figure 8.** Total charge-pickup cross section as a function of the projectile energy per nucleon – open triangles $^{197}$Au + $^{1}$H [11], full dot $^{197}$Au + $^{1}$H [25], full square $^{208}$Pb + $^{1}$H from the present work and open dots $^{197}$Au + $^{1}$H [9]. The data from Refs. [9] and [11] were extracted from measurements performed with $CH_2$ and C targets.

The partial charge-pickup cross sections are shown in Figure 9 as a function of the difference between the mass number of the resulting fragment and the mass number of the projectile. We have compared the data from the reaction $^{208}$Pb + $^{1}$H measured in the present work with data from the following reactions: $^{208}$Pb + Cu at 1 $A$ GeV [27], $^{197}$Au + $^{1}$H at 0.8 $A$ GeV [25], $^{129}$Xe + $^{27}$Al at 0.79 $A$ GeV [13] and $^{86}$Kr + $^{9}$Be at 0.5 $A$ GeV [14]. As the data on $^{208}$Pb + $^{2}$H and $^{208}$Pb + "Ti" from the present work are already shown in Figure 4, they are not included in Figure 9.

From previous works, where the influence of the size of the projectile nucleus on the total charge-pickup cross section was studied, it became clear that the total cross section is decreasing with decreasing projectile mass [10, 12, 38]. Firstly, it is expected that the primary production of charge-pickup prefragments is decreasing [13], and secondly, the lower Coulomb barriers in lighter nuclei and shorter distance from the evaporation corridor [28] lead to larger proton-evaporation probabilities during the decay of excited prefragments, thus resulting in lower cross sections for the survival of $Z_p + 1$ prefragments. From Figure 9 one can see that the proton evaporation depletes the neutron-deficient side of the isotopic distribution for the lighter elements while the most neutron-rich side is not influenced. This is the obvious result of the fact that the proton evaporation is becoming more competitive with increasing number of evaporated neutrons, especially in the case of lighter elements. In the case of heavy nuclei, the neutron-deficient side is depleted by both proton evaporation and fission.

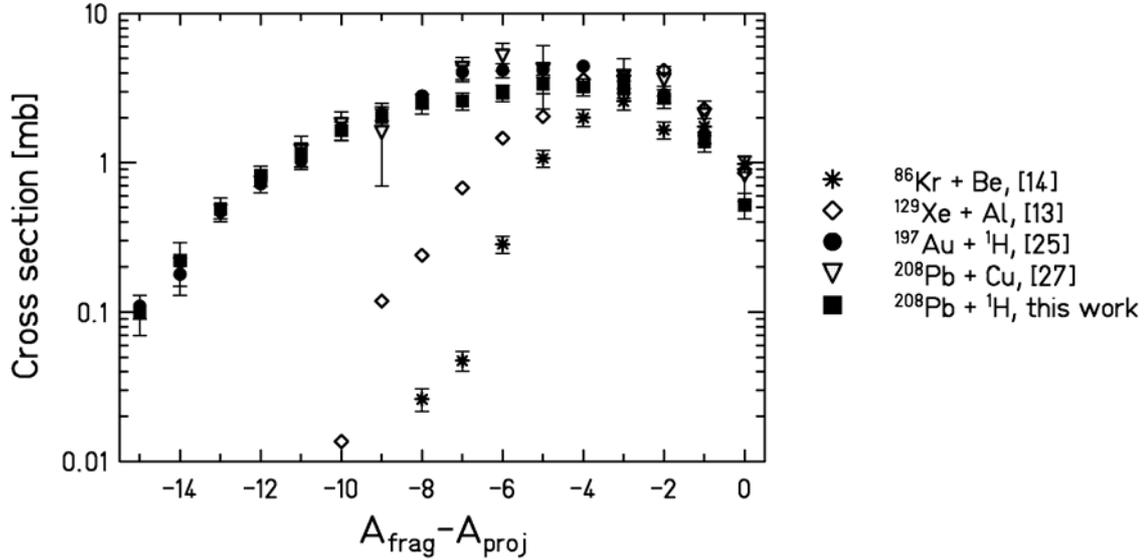

**Figure 9.** Comparison between the partial charge-pickup cross sections from the present work and available data. Data are shown as the function of the difference between the mass numbers of the resulting fragment and of the projectile for the following reactions: $^{208}$Pb + $^{1}$H at 1 $A$ GeV from the present work (full squares), $^{208}$Pb + Cu at 1 $A$ GeV [27] (open triangles), $^{197}$Au + $^{1}$H at 0.8 $A$ GeV [25] (full dots), $^{129}$Xe + $^{27}$Al at 0.79 $A$ GeV [13] (open diamonds) and $^{86}$Kr + $^{9}$Be at 0.5 $A$ GeV [14] (stars). All these measurements were performed at GSI. As the data on $^{208}$Pb + $^{2}$H and $^{208}$Pb + "Ti" from the present work are already shown in Figure 4, they are not repeated here.

Slightly higher isotopic cross sections for bismuth production in the $^{208}$Pb + Cu reaction [27] in Figure 9 compared to the $^{208}$Pb + $^{1}$H reaction from the present work are the consequence of heavier target mass as already discussed in section 3.1. Again it is seen that only few isotopes are influenced by the increase in target mass.

Comparing the data from the reaction $^{197}$Au + $^{1}$H at 0.8 $A$ GeV [25] with the data from the $^{208}$Pb + $^{1}$H reaction at 1 $A$ GeV one can see that for the isotopes having 3 to 7 nucleons less than the corresponding projectile, $^{190-194}$Hg from $^{197}$Au + $^{1}$H [25] and $^{201-205}$Bi from $^{208}$Pb + $^{1}$H, the cross sections for mercury production are higher than the ones for bismuth production. All other isotopes in these two reactions are produced with almost the same cross sections. On the other hand, Figure 8 shows that the total cross sections from these two reactions are following the behaviour of other data as a function of projectile energy. Therefore, we attribute this difference to the different beam energies used in these two reactions. Similar behaviour can be seen in the data reported by Gloris et al. [36]. Using the method of γ spectroscopy, they have measured the partial cross sections for the production of $^{204-207}$Bi in the reaction p + $^{nat}$Pb at several proton-beam energies. These data are shown in Figure 10. From this figure, no change in the charge-pickup cross sections for these heaviest products at the two highest proton-beam energies is evident. For the lowest energy in Figure 10, although the error bars are rather large, there is an indication for the increase of the production cross sections for $^{204}$Bi and $^{205}$Bi. These isotopes correspond to those in Figure 9 for which the difference in the partial charge-pickup production cross sections in the reactions $^{208}$Pb + $^{1}$H and $^{197}$Au + $^{1}$H starts to be visible.

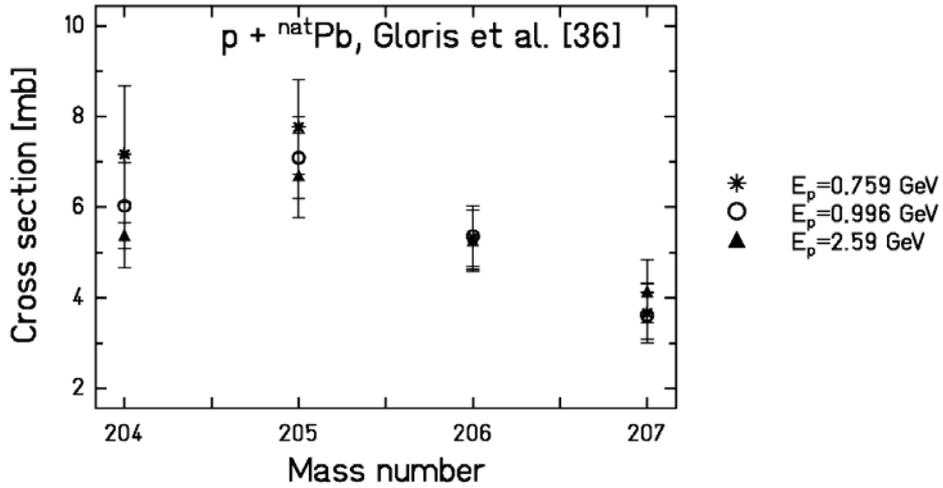

**Figure 10.** Partial charge-pickup cross sections for $^{204-207}$Bi from the reaction p + $^{nat}$Pb at different proton-beam energies. The data are taken from reference [36].

In Table 4, a part of the isotopic distribution from the reaction $^{208}$Pb + $^1$H measured in the present work is compared with γ-spectroscopy data from p + $^{208}$Pb reaction at a proton-beam energy of 1 GeV [37]. Data from Ref. [36] are not included in Table 4, as they were measured with natural lead target, and thus different lead isotopes contribute to the production of a given bismuth isotope. The data from Ref. [37] are higher than those measured in the present work by the factor of ~2. The origin of this difference is not clear to us.

**Table 4.** Partial bismuth production cross section measured in the present work (given are total uncertainties) and compared with γ-spectroscopy data from p + $^{208}$Pb [37] at a proton-beam energy of 1 GeV.

| | σ [mb] | | | |
| --- | --- | --- | --- | --- |
| | $^{204}$Bi | $^{205}$Bi | $^{206}$Bi | $^{207}$Bi |
| Present work | 3.22 ± 0.41 | 3.11 ± 0.40 | 2.70 ± 0.39 | 1.38 ± 0.20 |
| p + $^{208}$Pb [37] | 4.60 ± 0.29 | 6.20 ± 0.40 | 5.29 ± 0.80 | 4.84 ± 0.39 |

## 5. Comparison with model calculations

At relativistic energies, a charge-exchange reaction is conveniently described as a two-stage process [13]: In the first, faster, stage interactions between target and projectile nucleons create excited prefragments. This stage can be described by an intra-nuclear cascade model, where one follows a sequence of independent two-body nucleon-nucleon collisions. In the second, slower, stage the excited prefragments decay, and the competition between neutron evaporation, charged-particle evaporation and fission determines the number of final, observed, fragments.

To compare the data from the present work with calculations, we have used two different intra-nuclear cascade codes: ISABEL [40] and the latest version of the Liege code INCL4 [41], both coupled to the same evaporation-fission code ABLA [42] developed at GSI.

In the following, we will compare the calculated production cross sections and velocity distributions with the experimental results.

**5.1 Production cross sections**
In Table 2, the total calculated charge-pickup cross sections are compared with the measured cross sections. For the reaction $^{208}$Pb + "Ti" the calculations were performed only with ISABEL, because in the present version of INCL4 the heaviest target that can be used in calculations is $^{4}$He. The calculations give slightly higher total charge-pickup cross sections, but the values are still in the range given by the error bars of the measured cross sections.

In order to compare the model predictions with the experimental results in more detail, Figure 11 shows the calculated and the measured partial charge-pickup cross sections for bismuth production in the three reactions considered in the present work.

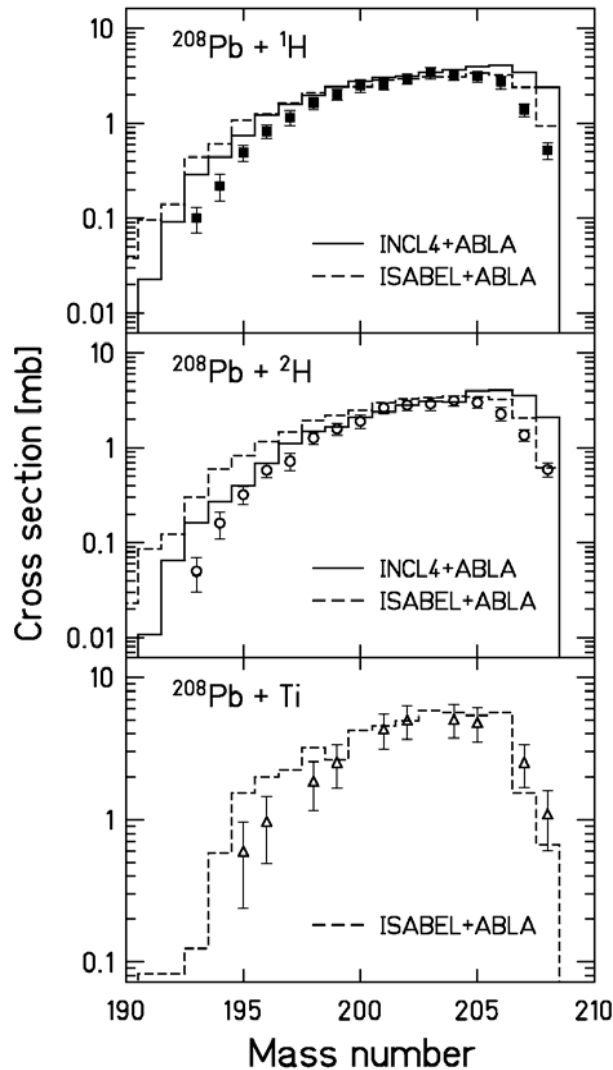

**Figure 11.** Partial cross sections for bismuth production – Comparison between experimental data, full squares: $^{208}$Pb + $^{1}$H, open dots: $^{208}$Pb + $^{2}$H, and open triangles: $^{208}$Pb + "Ti", with different model calculations: full line – INCL4 [41] + ABLA [42] and dashed line – ISABEL [40] + ABLA. Error bars represent total (systematic plus statistical) uncertainties.

In the case of $^{208}$Pb + $^{1}$H and $^{208}$Pb + $^{2}$H, the slightly higher values of the calculated total charge-pickup cross sections are reflected in wider isotopic distributions compared to the experimental ones. Both calculations over-predict the neutron-rich side with ISABEL giving better agreement with the experimental data than INCL4. The neutron-rich bismuth isotopes originate from prefragments having low excitation energy, and the problem of reproducing the low-excitation energy events was already noticed by Boudard et al. [41]. As the only quantum restriction considered is Pauli blocking, other quantum effects not considered in the model could be important in the cases where the low-excited prefragments are created through a few elementary nucleon-nucleon collisions [41]. On the other hand, ISABEL over-predicts the cross sections for the production of the most neutron-deficient bismuth isotopes, while INCL4 gives a better description of this part of the distribution. A possible origin of the difference in the description of the neutron-deficient side with these two models could be connected with the induced angular momenta. The root-mean-square of the angular momentum given by INCL4 is almost two times higher than the one given by ISABEL. The prefragments with higher angular momenta have higher probability for fission, and, as a consequence, the neutron-deficient side of the isotopic distribution is less populated.

For the $^{208}$Pb + "Ti" reaction, the calculation was performed assuming a pure titanium target, although the real target consists of several different materials as shown in Table 1. This simplification is justified by the fact that titanium represents an average of the present target layers and that also, as seen in Section 3, the charge-pickup cross sections vary slowly with the mass of the target nucleus. The total charge-pickup cross sections as well as the shape of the isotopic distributions are reproduced with the ISABEL calculations in a very satisfactory way, see Table 2 and Figure 11.

At the end of this paragraph, we would like to make few comments on the Pauli principle, a quantum prescription added to the semi-classical description of the projectile-nucleus interaction in INCL4. An obvious failure of INCL4 is the large over-prediction of the production cross section of $^{208}$Bi, which is basically produced in one nucleon-nucleon collision. Since free nucleon-nucleon cross sections are realistically described in the intra-nuclear cascade approach, these data carry valuable information on in-medium effects.

For each nucleon-nucleon interaction in INCL4, it is checked statistically that after a collision the nucleons will find a reasonable quantum place in the phase-space of the nucleus. Since all nucleons are stochastically positioned inside the nucleus, we do not have a compact medium when a collision occurs, even in the ground state. On the first collision, the target nucleon gains some energy but sometimes can find a free quantum cell below the Fermi level. This is questionable for a nucleus in the ground state but is not impossible if we have in mind a smooth occupation around the Fermi level allowed by the configuration mixing and the finite nucleus temperature. After several nucleon-nucleon collisions warming up the nucleus and ejecting nucleons, this mechanism takes into account the more and more dilute nuclear matter. Another possibility is to reject at least the first collision when a final nucleon is produced below the Fermi energy (strict Fermi blocking on the first collision). This implementation has been discussed and tested below 0.2 GeV [43] and gives reasonable improvements of the neutron and proton spectra, especially for their high energy part in direct kinematics. In the present case, inclusion of the strict Fermi blocking affects mainly the $^{208}$Bi production, which is reduced by a factor ~2 as compared to the standard calculations with statistic Fermi blocking, see Table 5.

**Table 5.** Calculated production cross sections for $^{208}$Bi and $^{207}$Bi in the reactions of $^{208}$Pb + $^{1}$H, $^{2}$H, with and without taking into account the strict Fermi blocking on the first collision.

|  | $^{208}$Pb + $^{1}$H | | $^{208}$Pb + $^{2}$H | |
|---|---|---|---|---|
|  | $^{208}$Bi | $^{207}$Bi | $^{208}$Bi | $^{207}$Bi |
| Strict Fermi blocking | 1.35 mb | 3.17 mb | 1.23 mb | 3.20 mb |
| Statistic Fermi blocking | 2.38 mb | 3.43 mb | 2.12 mb | 3.54 mb |

The cross section for the production of $^{207}$Bi is lowered by ~10%, and the total charge-pickup cross section reduced by ~6.5%. The cross sections for the production of other nuclides are not influenced by this effect. This substantial improvement driven by the present data will be inserted in the cascade code after some further tests.

**5.2 Velocity distributions**

The velocity distributions of bismuth isotopes produced in $^{208}$Pb + $^{1}$H, $^{2}$H were calculated using the INCL4 code coupled with ABLA. The results of the calculations, for the same bismuth isotopes considered in Figure 6, are shown in Figure 12. The calculated velocity distributions are normalised to the corresponding calculated cross sections.

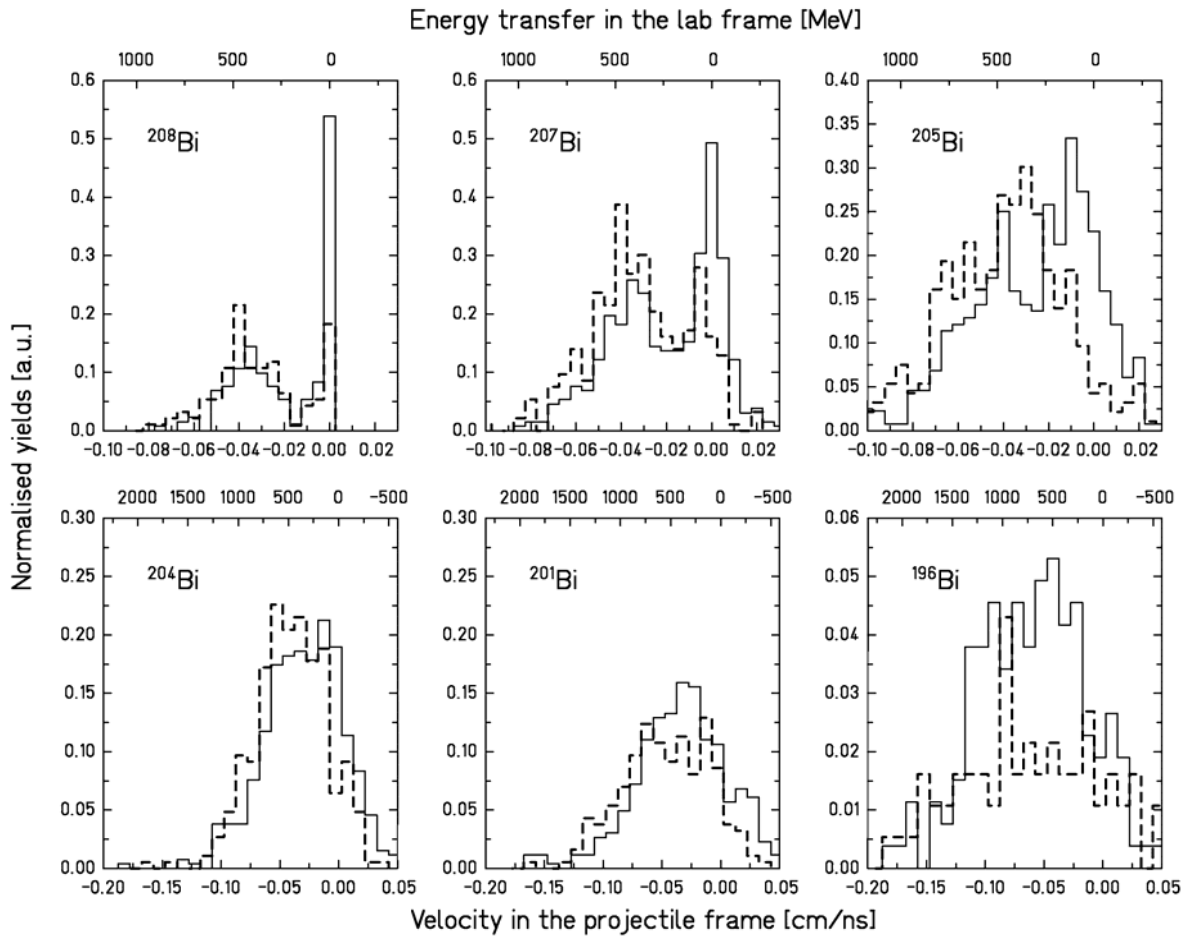

**Figure 12** Calculated velocity distributions of several bismuth isotopes produced in the interaction of 1 *A* GeV lead with the proton- (full line) and the deuteron- (dashed line). The velocity distributions are normalised to the corresponding calculated production cross sections. The calculations were performed with INCL4 + ABLA. The upper *x*-axis shows the energy transfer in the laboratory frame.

The general behaviour of the calculated velocity distributions is the same as for the experimental distributions shown in Figure 6.

The calculated velocity distributions are in a satisfactory agreement with the measured distributions. Also in the calculated distributions of $^{207-204}$Bi isotopes there is a hint of a stronger contribution of the quasi-elastic component in the proton target as compared to the deuteron target. In the case of $^{208}$Bi, the calculated velocity distributions show the same qualitative behaviour as in the case of the measured distributions – going from the proton to the deuteron target the quasi-elastic component is decreasing and the contribution from the Δ-excitation is increasing. On the other hand, there are quantitative differences between the calculated and the measured velocity distributions for this isotope. A 2.0 (2.2) times weaker quasi-elastic component and a 1.5 (1.3) times stronger contribution from the Δ-excitation in the deuteron target as compared to the proton target are obtained from calculations with statistic (strict) Fermi blocking, while from the measured distributions these numbers are 1.4 and 1.7, respectively. This discrepancy could give valuable information on some possible shortcomings in the theoretical description of the pure charge-exchange channel in the code.

## 6. Conclusion

Isotopically resolved charge-pickup cross sections and velocity distributions have been measured in the reactions of 1 $A$ GeV $^{208}$Pb with proton, deuteron and titanium target.

The total and partial charge-pickup cross sections in the reactions $^{208}$Pb + $^{1}$H and $^{208}$Pb + $^{2}$H are measured to be the same in the limits of the error bars. A weak increase in the total charge-pickup cross section is seen in the reaction of $^{208}$Pb with the titanium target.

The precise measurements of the velocity distributions of fully identified residues has been shown to be a powerful tool for disentangling different reaction mechanisms. The contributions from quasi-elastic scattering and Δ-resonance excitation to the production of the heaviest bismuth isotopes are clearly seen in the velocity distributions measured in the present work (Figure 6). With decreasing mass of the residues, the quasi-elastic component is disappearing, and for the lightest bismuth isotopes only the contribution from the Δ-resonance is present. Dying-out of the quasi-elastic component is seen to be faster for the deuteron and "Ti" target as compared to the proton target.

The total and partial charge-pickup cross sections from these three reactions are compared with other existing data. The data from the present work follow nicely the general behaviour with projectile energy and target size.

Being sensitive to the nucleonic aspects of relativistic heavy-ion collisions, data on charge-pickup reactions are an important test for any microscopic model on nucleon-nucleon interactions. In order to gain some understanding of this process, we have compared the data from the present work with the predictions of two intra-nuclear cascade models, INCL4 and ISABEL, coupled to the same evaporation-fission model, ABLA. Both models reproduce quite satisfactorily the measured total charge-pickup cross sections, while the widths of calculated isotopic distributions are larger than the experimental ones for all three targets. The velocity distributions of the final residues were also calculated and generally were found to be in rather good agreement with the measured distributions. In case of INCL4, it was shown that the experimental data can help to improve the treatment of the Pauli blocking in the code.

## Acknowledgments

We would like to thank to T. Hennino, B. Ramstein, M. Roy-Stéphan and K. Sümmerer for fruitful discussions during the course of this work. We thank also Prof. J. Cugnon for clarifying remarks on the INCL model.

This work was supported by the European commission in the frame of the HINDAS project under the contract number ERBCHBCT940717.